\documentclass[aps,pre,eqsecnum,amsmath,amssymb,showkeys,showpacs,twocolumn]{revtex4}
\usepackage[]{graphicx,color}
\graphicspath{{./}}

\renewcommand{\vec}[1]{{\bf #1}}
\newcommand{\Nabla}{\mbox{\bf\boldmath $\nabla$}}

\begin{document}

\title{
End-wall effects on the transition
between Taylor vortices and spiral vortices
}

\author{
	Sebastian Altmeyer$^1$, 
	Christian Hoffmann$^1$, 
	Matti Heise$^2$, 
	Alexander Pinter$^1$,
	Manfred L\"ucke$^1$, and 
	Gerd Pfister$^2$
}

\affiliation{
	$^1$Institut f{\"u}r Theoretische Physik, Universit{\"a}t des Saarlandes, D-66123 Saarbr{\"u}cken, Germany\\
	$^2$Institut f{\"u}r Experimentelle und Angewandte Physik, Universit{\"a}t Kiel, D-24098 Kiel, Germany
}

\pacs{
	47.20.Ky,
	47.32.cf,
	47.54.-r
}

\keywords{
	Taylor vortices,
	Spiral vortices, 
	Wavy structures,
	Secondary bifurcations
}

\begin{abstract}
We present numerical simulations as well as experimental results concerning
transitions between Taylor vortices and spiral vortices in the
Taylor-Couette system with rigid, non-rotating end-walls in axial direction.
As in the axial periodic case, these transitions are performed by wavy
structures appearing via a secondary bifurcation out of Taylor vortices and
spirals, respectively. But in the presence of rigid lids, {\em pure} spiral
solutions do not occur but are substituted by primary bifurcating, stable
wavy spiral structures (wSPI). Similarly to the periodic system, we found a
transition from Taylor vortices to wSPI mediated by so called wavy Taylor
vortices (wTVF) and, on the other hand, a transition from wSPI to TVF
triggered by a propagating defect. We furthermore observed and investigated
the primary bifurcation of wSPI out of basic Ekman flow.
\end{abstract}

\maketitle

\section{Introduction}

\begin{figure}
  \begin{center}
    \includegraphics[clip,width=0.95\linewidth]{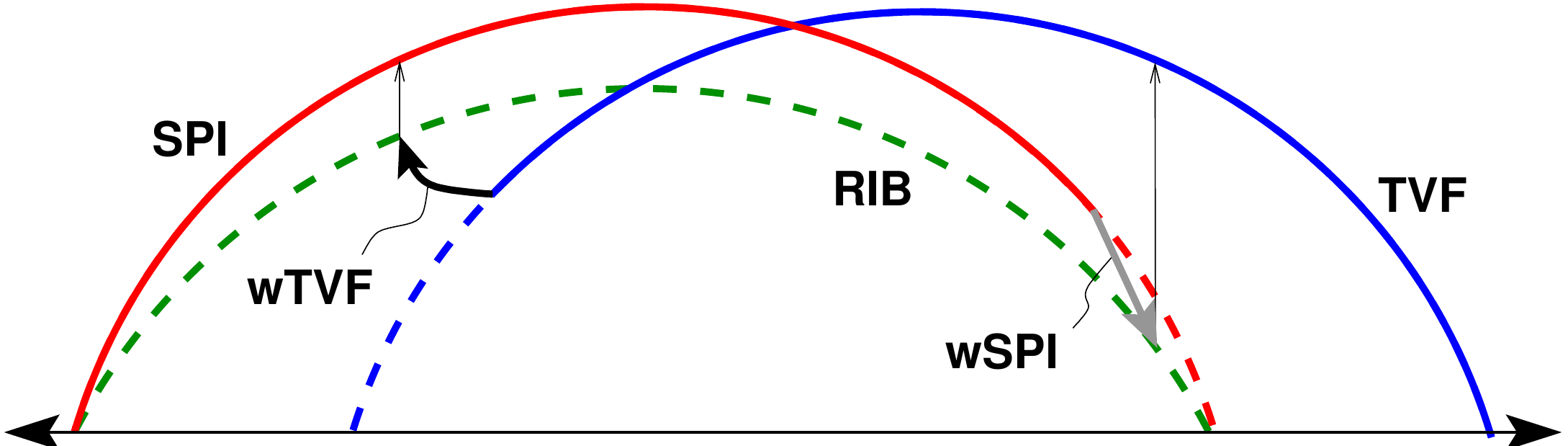}
    \caption{
    \label{fig01}
    (Color online) Schematic bifurcation diagram for a suitably
    chosen control parameter.  Stable (unstable) solutions are displayed as
    solid (dashed) lines.  Thin arrows indicate the transients corresponding
    to the 'jump' bifurcation mentioned in \cite{GSS88}.  
    }
  \end{center}
\end{figure}  

The interaction between Taylor vortex flow (TVF), spirals (SPI), and a
variety of different wavy solutions was investigated in numerous
publications \cite{GSS88,GL88,I86,J85,THS88,ALS85,AS00,AS02,CHSAML2009}.

Under periodic boundary conditions (pbc), toroidally closed TVF appears via
a primary stationary bifurcation out of the rotationally symmetric, axially
homogeneous basic circular Couette flow (CCF). Also the two axially symmetry
degenerated, oscillatory SPI states appear via primary bifurcations out of
CCF in a symmetric Hopf bifurcation together with the ribbon state. The
latter is typically unstable close to onset and can be seen as a non-linear
superposition of the two oppositely propagating spirals to an axially
standing wave. The stability of TVF and SPI at onset is regulated by the
order of their appearance upon increasing the inner cylinder's rotation
rate: the first (second) solution to bifurcate out of CCF is stable
(unstable). However, the second unstable solution becomes stable at larger
inner cylinder rotation. Which state bifurcates first depends on the outer
cylinder rotation rate \cite{T94}.

Besides parameter regions with mono-stability of TVF and SPI, one also
observes regions with bistability of both states \cite{HLP04}. When moving a
control parameter out of this region, one solution loses its stability and
the flow undergoes a transition to the remaining stable state, i.e., from
TVF to SPI or vice versa \cite{HLP04}. Bifurcation theoretical
considerations and symmetry arguments \cite{GSS88,GL88} as well as amplitude
expansion techniques in \cite{I86} and numerical simulations in
\cite{CHSAML2009} show that the stability is transfered from TVF to SPI via
secondarily bifurcating wavy Taylor vortices (wTVF) \cite{I86,J85} and vice
versa from SPI to TVF via secondarily bifurcating wavy spirals (wSPI). The
solution branch of stable TVF (SPI) is connected to unstable ribbons
\cite{CI94} via stable wTVF (wSPI). Then, a 'jump' bifurcation \cite{GSS88}
from the end of the stable wTVF (wSPI) branch leads to the stable SPI (TVF)
branch. This bifurcation behavior is schematically illustrated in
Fig.~\ref{fig01}.

Note that in the majority of publications, the wTVF solution branch has been
seen to return to the TVF branch or to undergo higher order bifurcations
\cite{J85,ALS85,AS00,AS02,GADCDL1983,KP1984} at larger driving.

On the other hand, rigid non-rotating lids at the axial ends (rigid boundary
conditions, rbc) induce rotational symmetric Ekman vortices even for
sub-critical driving \cite{T94}. This modifies the structure, stability and
bifurcation behavior of the different states.

One important difference between pbc and rbc is the absence of {\em pure}
SPI solutions for rbc, in particular, for moderate outer cylinder rotation,
i.e. within the left part of Fig.~\ref{fig01}. The rotational-symmetric
Ekman modes interact with the spiral modes leading to wSPI which bifurcate
primarily out of the basic Ekman state and play a similar role as the SPI
under pbc \cite{CHSAML2009}.

Furthermore, under rbc, wSPI loses its stability
for stronger inner cylinder rotation and we found traveling 
defects which trigger the transition from wSPI to stable TVF.

In \cite{MHCHJAAPGPML2008}, we described a situation where a similar defect
which separates domains of oppositely traveling spiral waves propagates
through the system and trigger a transition from right-handed to left-handed
spirals or vice versa. Thus, the transition from wSPI to TVF is a further
example for a transition which is triggered by a propagating defect.

This paper elucidates how TVF are transformed into SPI and vice versa under
the presence of Ekman induced disturbances, how stability is transferred
between the branches, and where and what kind of transients occur under rbc.
It is roughly divided into two parts corresponding to the transition from
TVF to wSPI via wTVF and the transition from wSPI to TVF via a propagating
defect. Structural dynamics, frequencies and wave number selection are
discussed and a comparison of the results obtained for rbc and pbc is made.

\section{System}

The Taylor-Couette system consists of a fluid of kinematic viscosity $\nu$
in the gap between two concentric, independently rotating cylinders (inner,
outer radius $r_{1,2}$; angular velocities $\Omega_{1,2}$; fixed radius
ratio $\eta=r_1/r_2=0.5$ and fixed length $\Gamma=12$ in units of the gap
width $d=r_2-r_1$) and non-rotating, rigid lids at the axial ends.

Cylindrical coordinates $r,\varphi,z$ are used to decompose the velocity
field into a radial component $u$, an azimuthal one $v$, and an axial one
$w$
\begin{eqnarray}
\vec{u}=u\,\vec{e}_r + v\,\vec{e}_\varphi + w\,\vec{e}_z.
\end{eqnarray}
The system is governed by the Navier-Stokes equations
\begin{eqnarray}
\partial_t \vec{u} = \Nabla^2 \vec{u} -
(\vec{u}\cdot \Nabla)\vec{u} - \Nabla p.
\end{eqnarray}
Here, lengths are scaled by the gap width $d$ and times by the radial
diffusion time $d^2/\nu$ for momentum across the gap and the pressure $p$ by
$\rho \nu^2/d^2$. The Reynolds numbers
\begin{eqnarray}
R_1=r_1\Omega_1 d/\nu,\quad R_2=r_2\Omega_2 d/\nu
\end{eqnarray}
enter into the boundary conditions for $v$. $R_1$ and
$R_2$ are just the reduced azimuthal velocities of the fluid at the cylinder
surfaces. Within this paper, we hold fixed $R_2=-100$.

For numerical simulations, we used the G1D3 code described in
\cite{HLP04,HL00}, i.e. a combination of a finite differences method in
radial $r$ and axial $z$ direction and a Galerkin expansion in $\varphi$
direction:

\begin{eqnarray} 
f(r,\varphi,z,t) &=&
  \sum_m
  f_{m}(r,z,t)\,e^{im\varphi}, \\
\nonumber
  && f \in \{u,v,w,p\}
\end{eqnarray}

In the experimental setup, the inner cylinder ($r_i=(12.50 \pm 0.01)$ mm) is
machined from stainless steel, while the outer cylinder ($r_2 = (25.00 \pm
0.01)$ mm) is made from optically polished glass. As fluid thermostatically
controlled silicone oil with a kinematic viscosity $\nu = 10.6$ cS is used.
At top and bottom the flow is confined by massive end-walls with a till
better than 0.03 mm at the outer diameter. The flow is visualized by
elliptical aluminum particles having a length of 80 $\mu$m. Flow
visualization measurements are performed by monitoring the system with a
CCD-camera in front of the cylinder recording the luminosity along a narrow
axial stripe. The spatio-temporal behavior of the flow is then represented
by successive stripes for each time step at a constant $\varphi$ position
leading to continuous space time plots.

\section{Transitions between TVF and wSPI}

As in the periodic system \cite{CHSAML2009}, we also found transitions
between the two primary bifurcating structures TVF and (w)SPI in the finite
length system. The main difference between both is that we do not observe
(neither in simulations nor in experiments) {\em pure} spirals for
$\Gamma=12$ systems, i.e. helical structures with the continuous symmetry
given by $f(r,\varphi,z,t)=f(r,kz+M\varphi-\omega t)$ for fixed $M$, $k$,
and $\omega$ (c.f.~\cite{HLP04}). In finite systems with non-rotating lids
generating Ekman vortices with an exponentially decaying amplitude in axial
direction, pure SPI are replaced by wSPI with a more complex mode spectrum
as described in \cite{CHSAML2009}. This is due to the interaction between
the spiral and the Ekman modes.  This section describes first the
bifurcation from TVF to wSPI via wTVF and after that the transition from
wSPI to TVF.

\subsection{Bifurcation from TVF to wSPI}

\begin{figure}  
  \begin{center}
    \includegraphics[clip,width=0.95\linewidth]{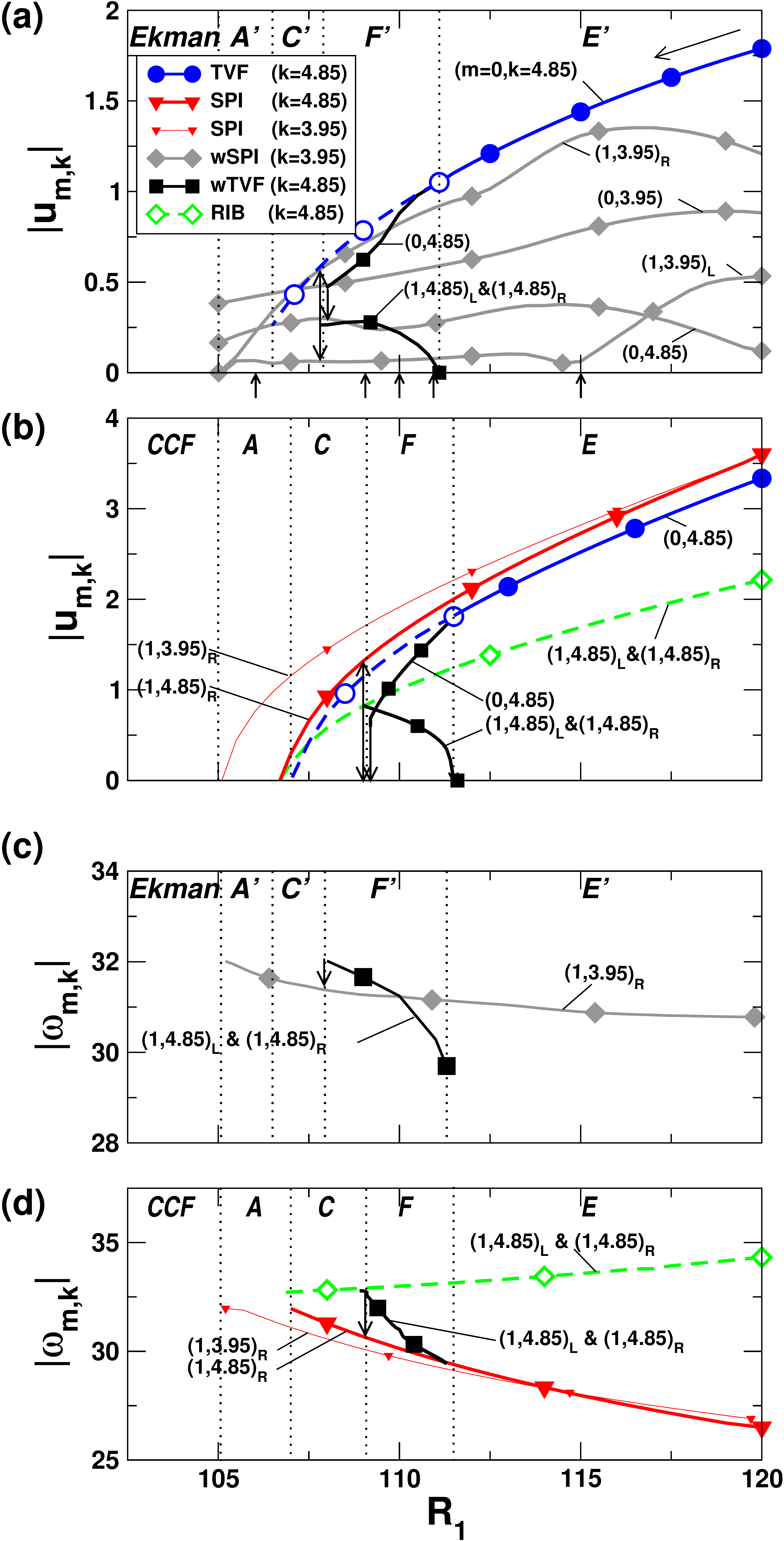}
    \caption{
    \label{fig02}
    (Color online) Numerically obtained bifurcation diagrams for different vortex structures TVF
    (blue, circles), SPI (red, triangles), wTVF (black, squares), wSPI
    (gray, diamonds), and ribbons (RIB, green, diamonds) versus $R_1$ for
    rbc (a,c) as well as for pbc (b,d). Curves with the same color, symbol,
    and line-width represent different modes of the same solution. Solid
    (dashed) lines with filled (open) symbols refer to stable (unstable)
    states. Shown are the significant radial flow field amplitude modes
    $|u_{m,k}|$ at mid-gap (a,b) and the corresponding frequencies
    $|\omega_{m,k}|$ (c,d).  The indices R and L correspond to right- and
    left-handed spiral modes.  The short arrows pointing to the abscissa in (a)
    denote the $R_1$ values of the snapshots in Fig.~\ref{fig03}, the long
    arrow in (a) indicates the direction for the transition TVF
    $\longrightarrow$ wTVF $\longrightarrow$ wSPI.  The sections labelled
    with letters correspond to different stability regions as listed in the
    table (c.f. Fig.~3 in \cite{CHSAML2009}):}
    \end{center}
      \parbox{.6\linewidth}{
        \begin{tabular}{lcccccccc}
            region & A & A' & C & C' & F & F' & E & E'  \\\hline\hline
            TVF    & - & -  & u & u  & u & u  & s & s   \\
            SPI    & s & -  & s & -  & s & -  & s & -   \\
            wSPI   & - & s  & - & s  & - & s  & - & s   \\
            wTVF   & - & -  & - & -  & s & s  & - & -   \\
            RIB    & u & -  & u & -  & u & -  & u & -   \\
          \end{tabular}
      }
      \parbox{.3\linewidth}{
          stable (s) \\
          unstable (u) \\
          nonexistent (-)
      } 
\end{figure}

Fig.~\ref{fig02} depicts the bifurcation branches for the interesting states
(w)SPI, (w)TVF, and ribbons. The different structures are distinguished by
symbols and line colors and characterized by amplitudes and frequencies of
their significant Fourier modes $(m,k)$ determined by the azimuthal wave
number $m$ and the axial wave number $k$ of the complete structure.  The
latter is included in the legend box. Solid (dashed) lines with filled
(open) symbols represent stable (unstable) solutions. Fig.~\ref{fig02}(a)
and (b) present mode amplitudes $|u_{m,k}|$ of the radial velocity field $u$
in the finite case (a) and the periodic case (b), respectively. (c) and (d)
show the corresponding frequencies $|\omega_{m,k}|$.

With our way of characterizing the flow by the combination of azimuthal and
axial Fourier modes, with the latter being obtained over the full axial
extension of the system, we do not capture, e.g. the Ekman induced axial
variation of $m=0$ modes.

We start our discussion of the bifurcation diagram in Fig.~\ref{fig02}(a) in
region E' with a stable $k=4.85$ TVF state which loses its stability in
region F' and C'. This is exactly the same behavior as under pbc (b), except
that the stability thresholds are slightly shifted, that the unstable TVF
branch bifurcates out of the $k=3.95$ Ekman state in (a) instead of the CCF
as in (b), and -- more importantly -- that the final state is a wSPI and not a
pure SPI. We omit the Ekman branch itself in the figure due to visibility
reasons.

In F' (and also in F), TVF becomes unstable against wTVF maintaining the
same wave number ($k=4.85$). Note that in (b), the wave number is determined
by the predefined periodicity length. At the right border of C (C'), wTVF
undergoes a transition to the remaining stable SPI (wSPI) solution with
$k=4.85$ ($k=3.95$), i.e., the wave number changes during this transition --
c.f.~\cite{CHSAML2009} for a detailed description of the stability
properties, the bifurcation behavior, and the structure of wTVF and wSPI.

We added the corresponding $k=3.95$ SPI solution branch in (b) in order to
emphasize the identical onsets of $k=3.95$ SPI (b) and $k=3.95$ wSPI (a). As
the SPI solution in (b), also the wSPI solution in (a) is stable within the
whole parameter range displayed here. The transition from wTVF to SPI (wSPI)
includes an unstable transient ribbon state, but we did not try to stabilize
this state for rbc.

Generally speaking, there are three major aspects concerning the finite and
the periodic system: (i) wSPI in (a) play quite the same role as the pure
SPI in (b) -- we indicated this by the prime at the labels A,C,E,F
distinguishing the different stability regions
(c.f.~\cite{HLP05,CHSAML2009}). (ii) for transition TVF $\longrightarrow$
wSPI, the finite $\Gamma=12$ system selects the same wave number $k=4.85$
for {\em all} toroidally closed structures (TVF, wTVF) and $k=3.95$ for the
helical solution (wSPI). Thus, transitions from TVF to wSPI are generally
accompanied by a change in the wave number. (iii) finite boundary conditions
superimpose rotational symmetric disturbances. Therefore, all thresholds
(dotted vertical lines in Fig.~\ref{fig02}(a,b)) of solutions with
rotational or toroidal symmetry are shifted towards lower $R_1$ compared to
the respective thresholds in the periodic system. The wTVF onsets E-F and
E'-F' coincide very well in both cases.

In Fig.~\ref{fig02}(a), the dominant mode $(0,4.85)$ of the unstable
$k=4.85$ TVF branch ends up in the $(0,4.85)$ Ekman mode at A'-C' which is a
sub-dominant mode in the $k=3.95$ Ekman state. Therefore, we included the
mode $(0,4.85)$ which is a higher Fourier mode of the $k=3.95$ wSPI state.

Note that due to the absence of symmetry breaking effects like axial
through-flow, right-handed and left-handed spiral solutions are equivalent
\cite{HLP04,HLP05} and therefore simply indicated by (w)SPI.

\subsubsection{Frequencies}

Fig.~\ref{fig02}(c) and (d) provide the frequencies $|\omega_{m,k}|$ of the
corresponding mode amplitudes in (a) and (b).  We omit those 
frequencies of TVF and wTVF which are zero.

SPI -- for pbc (d), spirals and ribbons grow via a primary Hopf bifurcation
with a common frequency out of CCF. The difference between the spiral
frequencies for pbc (d) and rbc (c) is a consequence of the Reynolds-stress
driven (intrinsic) axial net flow which is directed oppositely to the spiral
propagation. In finite systems, this net flow is suppressed by impermeable
lids which leads to a shift in the axial phase velocity and thereby also in
the frequency \cite{HLP04}. This effect can also be seen for the wSPI in (c)
and the SPI in (d): whereas the frequencies of both spirals ($k=4.85$ and
$k=3.95$) are nearly identical at E-F, the frequencies of the $k=3.95$ SPI
at E-F and the $k=3.95$ wSPI at E'-F' differ.

wTVF -- On the other hand, the wTVF frequencies at the bifurcation
thresholds E-F and E'-F' are almost identical. Here, the intrinsic net flow
of the $(1,4.85)_L$ mode is compensated by that of the $(1,4.85)_R$ mode in
both cases (c) and (d).  Moreover, the variation of wTVF and SPI frequencies
in region F differs significantly.

Since wTVF is a time-periodic rotating state that does not propagate
axially, all mode frequencies are either zero ($\omega_{0,4.85}=0$) or
multiples of $\omega_{1,4.85}$. So, the dynamics of wTVF is rather simple
while the spatial structure is more complex.

\subsubsection{Spatio-temporal behavior}

\begin{figure}
  \begin{center}
    \includegraphics[clip,width=0.95\linewidth]{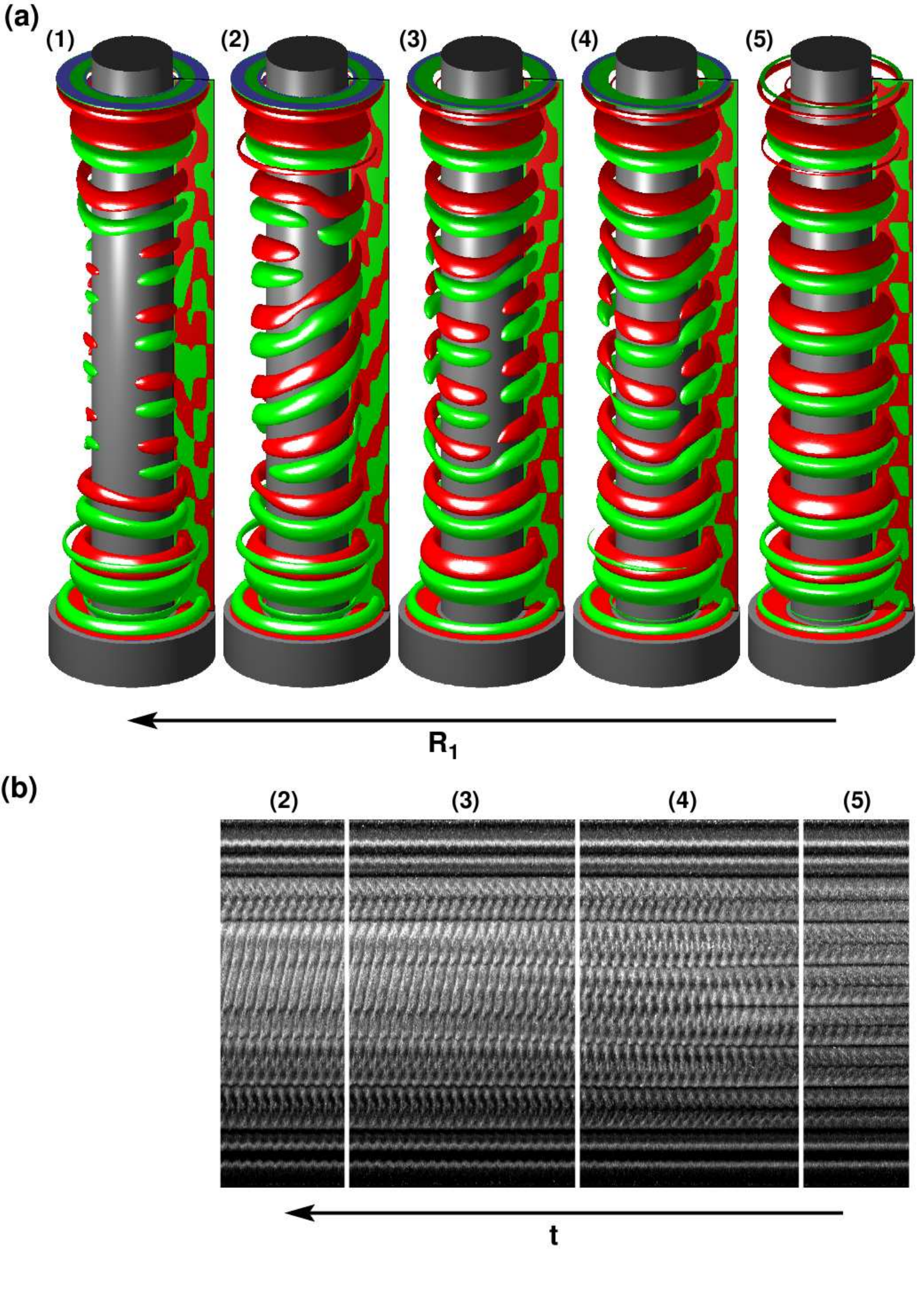} 
    \caption{
    \label{fig03}
    (Color online) (a) Numerical simulations: snapshots of isosurfaces of the azimuthal vorticity
    $\partial_zu-\partial_rw=\pm40$ (red: +40, green: -40) at five different
    $R_1$ values marked by arrows in Fig.~\ref{fig02}(a) during the
    transition TVF $\longrightarrow$ wTVF $\longrightarrow$ wSPI (from right
    to left).  Red (green) coloring on the additional $\varphi=$const. plane
    denotes positive (negative) vorticity.  We use $4\pi$ cylinders in (a)
    in order to present the whole structure in one single 3-dimensional
    plot.
    (b) Experimentally obtained spatio-temporal flow visualization of the
    transition which was triggered by an initial jump from $R_1=115$ to 109.
    The plots cover the complete system length of
    $\Gamma=12$. 
  }
  \end{center}
\end{figure}

In order to elucidate the  different states arising during the transition
TVF (snapshot \#5) $\longrightarrow$ wTVF (\#4, \#3) $\longrightarrow$ wSPI
(\#2, \#1), Fig.~\ref{fig03}(a) gives isosurfaces of the azimuthal
vorticity $\partial_zu-\partial_rw$ of stationary states at different $R_1$
values which are marked by short arrows pointing to the abscissa of 
Fig.~\ref{fig02}(a). 

The pure $k=4.85$ TVF state (\#5) becomes unstable against  toroidally
closed but axially modulated $k=4.85$ wTVF (\#4). Obviously, the modulation
strength increases towards mid-system where the Ekman influence is minimal.
As the $m\ne0$ mode contributions grow, the formerly rotational symmetric
structure becomes more and more deformed and the vorticity tubes narrow at a
certain $\varphi$ position (\#4). This means that the maximal vorticity
within the $(r,z)$ plane at this $\varphi$ position decreases with $R_1$ --
the vortex 'intensity' becomes weaker there. Note that this indentation of
the vortex tubes as well as the defect rotate with the whole structure.

Finally, the isosurfaces are completely constricted and separated (\#3). After
displacing the ends of the tubes, new connections are established and the
vorticity increases now to the final distribution in the $k=3.95$ wSPI (\#2).

The last snapshot, (\#1), depicts a situation in which the system is very
weakly supercritical and therefore, the Ekman vortices remain the dominant
structure.

Fig.~\ref{fig03}(b) presents the experimentally obtained spatio-temporal
behavior describing the dynamics of the different states after an initial
jump from $R_1=115$ to $109$ at the right border of the plot beginning with
a pure TVF state in (\#5) with $k\approx 4.8$ which then undergoes
(beginning at mid-height) a transition to wTVF with the same axial wave
number (\#4). After a transient (\#3) which corresponds to the 'jump'
bifurcation described above, wSPI with $k=3.6$ are finally established
(\#2). Note that we also verified in further experiments the stability and
stationarity of the TVF (\#5), wTVF (\#4), and wSPI (\#2) states for
suitable fixed control parameters $R_1$.

\subsubsection{Wave number selection}
\label{SEC:wave-change}

Due to the finite boundary conditions, the toroidally closed structures
(TVF, wTVF) can occur with discretely different axial wave numbers depending
on the initial conditions. We found at least three stable TVF states with 7
($k=3.83$), 8 ($k=4.85$), and 9 ($k=5.81$) vortex pairs in region E' (only
the 8 vortex pairs TVF state is presented in Fig.~\ref{fig02} and
\ref{fig03}). All of them undergo a transition to wTVF in region F' for
specific $R_1$ values without changing their respective wave number $k$.
Finally, all wTVF states 'jump' (accompanied by a change in $k$) to the
$k=3.95$ wSPI solution.

The experimentally ($k=4.53$ for (w)TVF and $k=4.03$ for wSPI) and the
numerically ($k=4.85$ for (w)TVF and $k=3.95$ for wSPI) obtained axial wave
numbers differ slightly. However, numerical simulations as well as 
experimental results exhibit the same jumps in $k$ during the
transition wTVF $\longrightarrow$ wSPI.

\subsection{Transition from wSPI to TVF}

\begin{figure}
  \begin{center}
    \includegraphics[clip,width=0.95\linewidth,height=.6\linewidth]{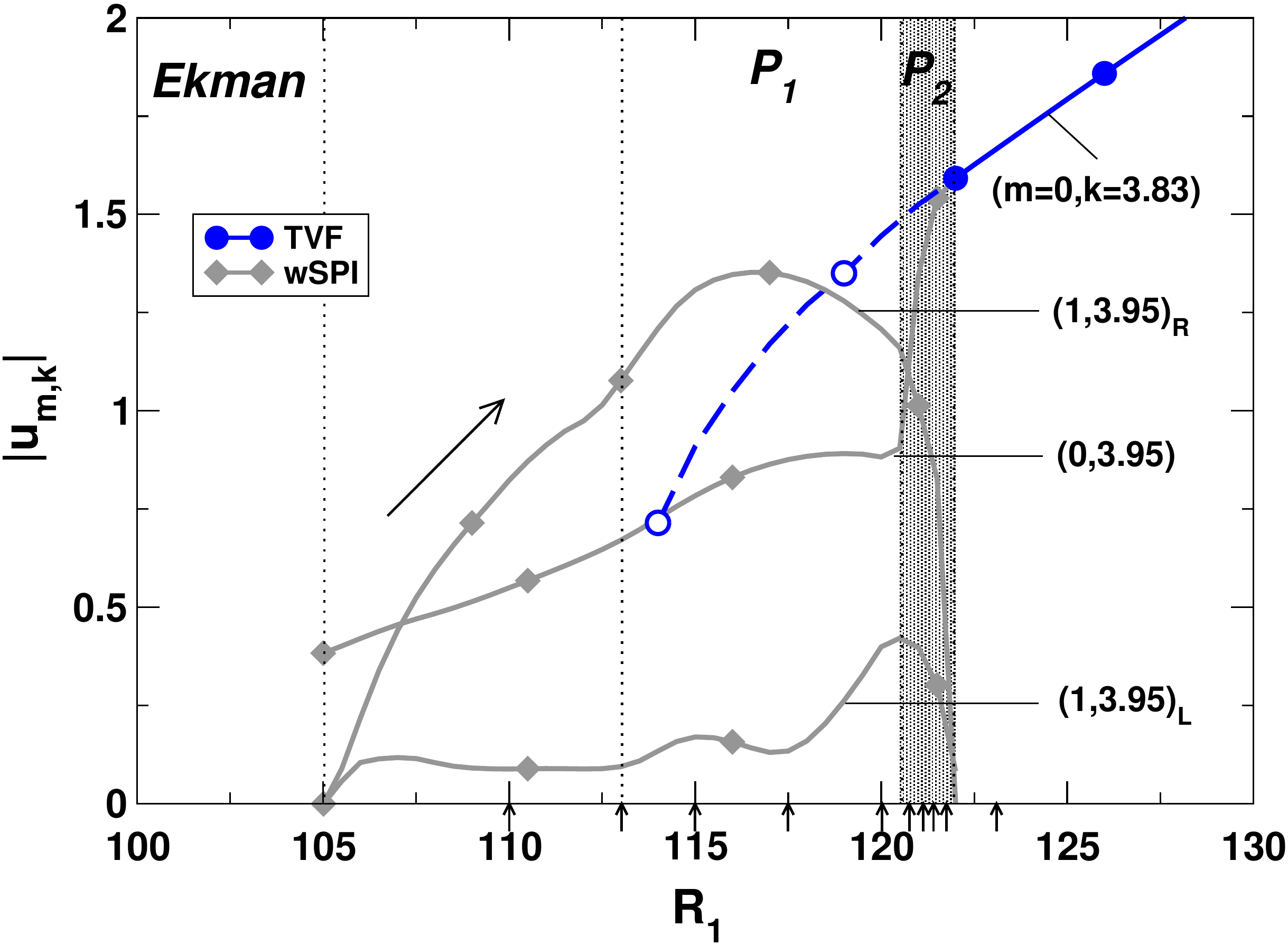}
    \caption{
    \label{fig04}    
    (Color online) Extension of Fig.~\ref{fig02}(a) -- see description
    there. Short arrows pointing to the abscissa identify the snapshots
    in Fig.~\ref{fig05}(a). The long arrow indicates the direction of the
    transition wSPI $\longrightarrow$ TVF.
    }
  \end{center}
\end{figure}

Under pbc, one finds the bifurcation sequence SPI $\longrightarrow$ wSPI
$\longrightarrow$ TVF. While the transition from wSPI to TVF is mediated by
a 'jump' bifurcation \cite{GSS88,CHSAML2009}, pure SPI, on the other hand,
could not be observed neither in rbc simulations nor in experiments.

As described above for rbc, wSPI bifurcates primarily out of the basic Ekman
flow as a stable solution for stronger counter-rotation. However, we found
wSPI to become unstable against TVF via an other kind of transition taking
place beyond region E' of Fig.~\ref{fig02}(a). This transition is mediated
by a propagating defect which separates the wSPI from a wTVF
regime pushing wSPI out and pulling wTVF through the bulk. Once the defect
crossed the whole bulk, the modulation amplitude of the wTVF vanishes and
pure TVF remains.

Fig.~\ref{fig04} gives a enhanced version of the bifurcation diagram in
Fig.~\ref{fig02}(a) with a slightly extended $R_1$ range covering the
transient wSPI $\longrightarrow$ TVF transition in the gray marked region
P$_2$. The short arrows mark the $R_1$ values for which isovorticity
snapshots are presented in Fig.~\ref{fig05}(a).

Starting with a pure wSPI solution at small $R_1$, an additional defect is
generated near the upper Ekman-spiral defect after increasing $R_1$ beyond
the left border of region P$_1$ and remains at its axial position for any
$R_1$ within the whole region P$_1$. At the right border of P$_1$, the
defect begins to propagate towards the other axial end. This is a transient
state which ends up in a pure $k=3.83$ TVF solution after the annihilation
of the propagating spiral-spiral defect at the lower Ekman-spiral defect. The gray
marked region P$_2$ gives the behavior of the amplitudes during this
transient from wSPI to TVF which occurs within the range $120<R_1<122$. Note
that the final TVF ($k=3.83$) is different from that discussed in
Fig.~\ref{fig02}(a) which has $k=4.85$. This TVF state is one of several
stable TVF states with different wave numbers (c.f.
Sec.~\ref{SEC:wave-change}).

This transition sequence agrees with experimental results as presented in
Fig.~\ref{fig05}(b) showing a spatio-temporal flow visualization of the
transition even after an initial jump from $R_1=107$ to 120 at the left
border of the plot.

\subsubsection{Spatio-temporal behavior}

\begin{figure*}[]
  \begin{center}
    \includegraphics[clip,width=0.95\linewidth]{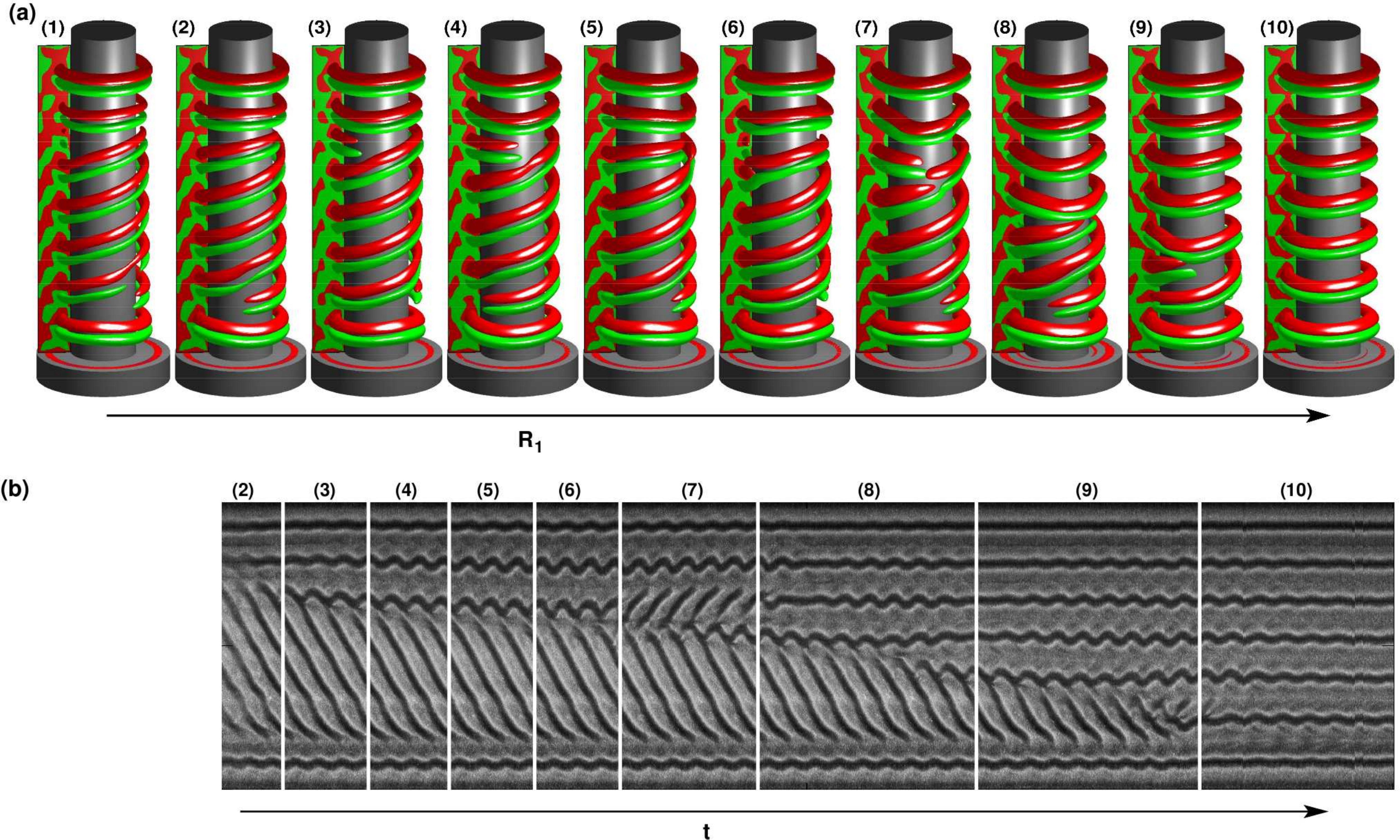}
    \caption{
    \label{fig05}
    (Color online) (a) Numerical simulations: snapshots of isosurfaces of the azimuthal vorticity
    $\partial_zu-\partial_rw=\pm40$ (red: +40, green: -40) of flow
    states at ten different $R_1$ values marked by short arrows pointing to the
    abscissa in Fig.~\ref{fig04} and visualizing the transition wSPI
    $\longrightarrow$ TVF while increasing $R_1$.  Red (green) coloring on
    the additional $\varphi=$const. planes denote positive (negative)
    vorticity.  $4\pi$ cylinders are used in order to present the whole
    structure in one single 3-dimensional plot.  Note that the propagating
    defect (\#5-\#9) is a transient state.
    (b) Experimentally obtained spatio-temporal flow visualization of the
    transition which was triggered by an initial jump from $R_1=107$ to 120.
    In axial direction, the plots cover the complete system length of
    $\Gamma=12$. 
    }
  \end{center}
\end{figure*}

The arrows in Fig.~\ref{fig04} mark the $R_1$ values of the ten snapshots of
Fig.~\ref{fig05} depicting the isosurfaces of the azimuthal vorticity (a)
as well as a spatio-temporal plot of the experimentally obtained velocity
field (b). Both sequences illustrate the structural changes during the
transition wSPI $\longrightarrow$ TVF mediated by a propagating defect.

Starting with a wSPI state (\#1) in Fig.~\ref{fig05} and increasing $R_1$,
the second upper closed vortex becomes wavy-like deformed in axial direction
(\#2-\#5) while the deformation rotates with the whole structure. The
modulation becomes stronger with increasing $R_1$ until a second defect
evolves out of the phase generating Ekman spiral defect (\#2). 

In (\#5), this defect detaches from the upper Ekman vortex and begins to
propagate (\#6)-(\#9), pushing the wSPI and pulling a wTVF state through the
bulk. Therefore, the localized wSPI region shrinks and the localized wTVF
domain grows. During this process, neither the wave number of wSPI nor that
of wTVF change significantly, because new wTVF vortices are generated
directly behind the propagating defect: comparing (\#7) and (\#8), one
observes that the vortex tubes become constricted and separated and after
displacing the ends of the tubes, new connections are established and a new
wTVF vortex is generated. Finally, the defect reaches the bottom end and
merges with the lower Ekman vortex (\#10) leaving behind a pure TVF state.

As described above, Fig.~\ref{fig05}(b) elucidates the spatio-temporal
behavior during the transition after an initial jump from $R_1=107$ to
$120$ at the left border of the plot. As in the numerical simulations, the
experimental setup realizes a transition from wSPI to TVF via a propagating
defect. Furthermore, the experimental wave numbers for wSPI $k=4.03$ and TVF
$k=3.84$ agree very well with the numerical wave numbers for wSPI $k=3.95$
and TVF $k=3.83$.

We'd like to stress that in the here presented experiments, the transitions
are initiated by an instantaneous jump (after preparing the initial state)
into the parameter regime where the final structure was expected to be
stable. The experimentally obtained spatio-temporal plots disclose a
sequence of transient structures which correspond to stationary structures
when quasi-statically driving the system by a $R_1$ ramp. We indicated the
corresponding structures by identical numbers of the snapshots in
Fig.~\ref{fig03} and \ref{fig05}.

\section{Summary}

\begin{figure}
  \begin{center}
    \includegraphics[clip,width=0.45\textwidth]{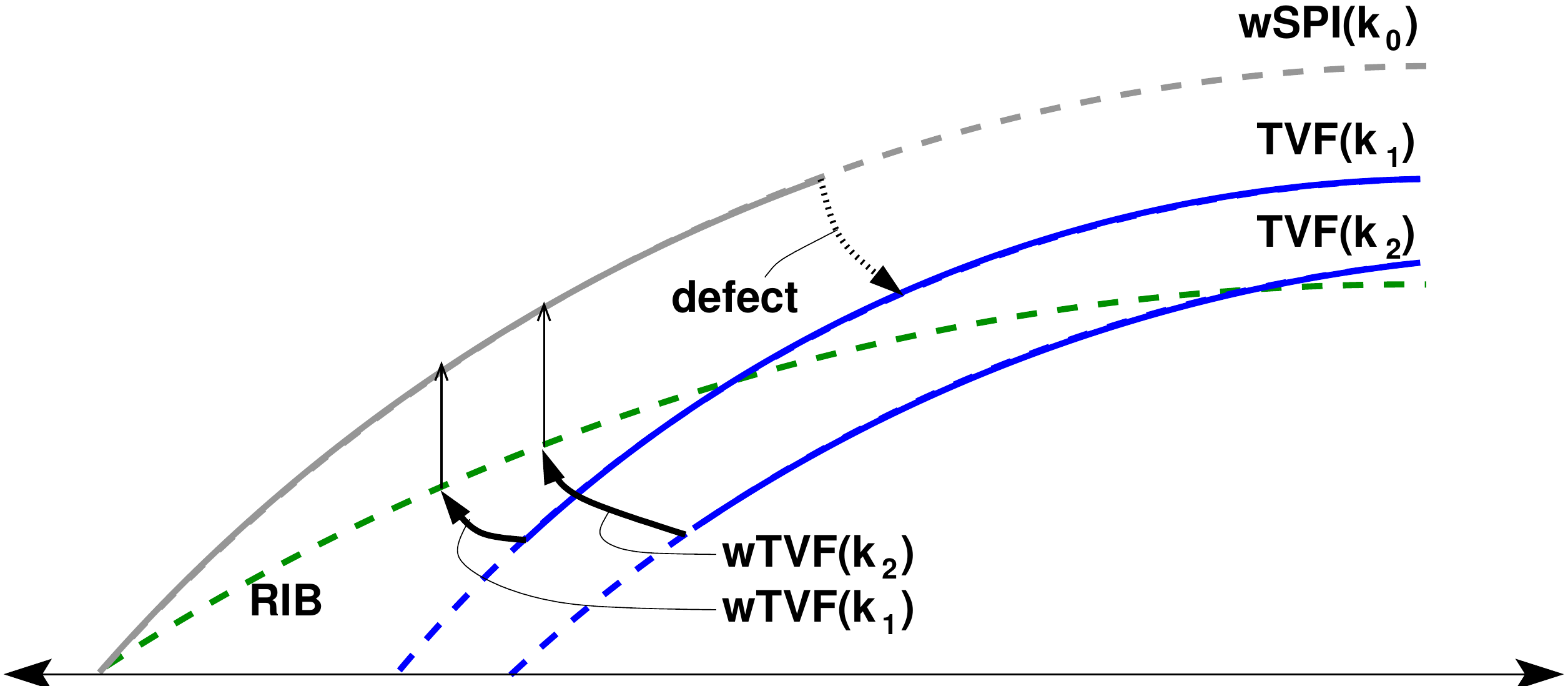}
    \caption{
    \label{fig06}
    Schematic bifurcation diagram for a suitable chosen
    control parameter containing the results of Fig.~\ref{fig02}(a) and
    Fig.~\ref{fig04} for the finite system (in contrast to Fig.~\ref{fig01}
    for the periodic system). Stable (unstable) solutions are displayed as
    solid (dashed) lines. Thin arrows indicate the transients corresponding
    to the 'jump' bifurcation. We included the ribbon branch as an
    intermediate unstable solution.
    }
  \end{center}
\end{figure}

We investigated the bifurcation behavior for the transition between
Taylor vortices and wavy spirals in a finite length Taylor-Couette system
with non-rotating, rigid lids. In contrast to periodic boundary conditions
where {\em pure} SPI solutions exist (even for outer cylinder at rest),
here, helical solutions occur as wavy structures due to the
admixture of Ekman induced $m=0$ mode components in the Fourier spectra. 

Under finite system geometry, we found a transition from TVF to wSPI via
wTVF which is analogue to the transition from TVF to SPI via wTVF in the
periodic system. wSPI (helical open vortices) are selected with a distinct
wave number $k_0$ whereas several bifurcation branches corresponding to TVF
and wTVF states (toroidally closed vortices) exist simultaneously and
multi-stably with different axial wave numbers (e.g. $k_1$ and $k_2$), i.e.
different numbers of vortex pairs (as schematically depicted in
Fig.~\ref{fig06}). The transitions from TVF to wSPI and vice versa are in
general accompanied by a change in $k$.

An other kind of transition performing the change from wSPI to TVF is
triggered via a propagating defect. This defect pushes the wSPI out of the
system and pulls wTVF and finally TVF into the bulk.

The coincidence of the wTVF frequency and the SPI frequency at the
bifurcation point disappears in the finite system.

\section*{Acknowledgement}
We thank the Deutsche Forschungsgemeinschaft for support.



\end{document}